\documentstyle[12pt]{article}
\begin{document}
\title{ 
A (2+1) dimensional integrable spin model: Geometrical and gauge 
equivalent counterpart, solitons and localized coherent structures}
\author{
R.
MYRZAKULOV$^{a,}$\thanks{rat@nlpub.ksisti.alma-ata.su 
and myrzakulov@hepi.academ.alma-ata.su} , 
  S. VIJAYALAKSHMI$^b$, \\ G. N. NUGMANOVA$^a$  and
M. LAKSHMANAN$^{b,}$\thanks{lakshman@cnld.rect.ernet.in} \\
$^a$High Energy Physics Institute, National Academy of Sciences, \\
480082,  Alma-Ata-82, Kazakstan\\ 
$^b$Centre for Nonlinear Dynamics, Department of Physics,          \\
Bharathidasan University, Tiruchirapalli 620 024, \\
India}

\date{}
\maketitle 
\begin{abstract}
A non-isospectral (2+1) dimensional integrable spin equation is 
investigated. It is shown that its geometrical and gauge equivalent 
counterparts is the (2+1) dimensional nonlinear Schr\"odinger equation 
introduced by Zakharov and studied recently by Strachan. Using a Hirota 
bilinearised form, line and curved soliton solutions are obtained. 
Using certain freedom (arbitrariness) in the solutions of the 
bilinearised equation, exponentially localized dromion-like     
solutions for the potential is found. Also, breaking soliton solutions 
(for the spin variables) of the shock wave type and algebraically localized 
nature are constructed.
\end{abstract}

\newpage
The nature of (1+1) dimensional integrable systems is now well understood [1].
On the  other hand, examples of (2+1) dimensional integrable equations 
solvable by Inverse Scattering Transform method (IST) are fewer in number and such
systems are being actively investigated from different points of view
at present [1,2]. An interesting subclass of integrable systems, useful both 
from the mathematical and physical points of view, is the set of integrable 
spin systems. Since  the identification of the first integrable spin 
model twenty years ago, namely the continuum isotropic  Heisenberg spin systems
[3,4], several  other integrable spin systems in (1+1) dimensions have 
been identified and investigated (see for example, refs.[5,6]) through 
geometrical and gauge equivalence concepts and IST method.

Again in (2+1) dimensions, only a small  number of integrable 
spin systems are known, among which the Ishimori equation (IE) is 
the most prominent one admitting different kinds of spin excitations 
such as solitons, vortices, dromions and so on[2]. Since the spin 
equations can also be considered as the nonrelativistic version of 
O(3) sigma models, which in turn have wide physical 
ramifications in (2+1) dimensions in problems such as high $T_c$ superconductors and quark 
confinement, construction of any integrable spin model                                                         in higher dimensions
assumes considerable significance. Moreover, since the spin systems have a 
natural connection with nonlinear Schr\"odinger family of equations in general, 
it is relevent from the soliton theory point of view to establish such 
connections in (2+1) dimensional cases also (as in the case of  IE and Davey
-Stewartson equation).
         
Recently a new family of integrable and nonintegrable (but admitting
exact solitary wave solutions) (N+1) dimensional (N = 1,2) classical spin 
models was proposed in refs[7-9]. One interesting nonlocal integrable spin model is
(the so called M-I equation)[7]

$$ \vec S_t = \{ \vec S \wedge \vec S_y + u \vec S \}_x, \eqno(1a) $$

$$      u_x = - \vec S \cdot \left( \vec S_x \wedge \vec S_y \right ). \eqno(1b)$$      
Here, subscripts stand for partial derivatives, $ \vec S= 
(S_1, S_2, S_3) $ and $ \vec S^2= S_3^2+r^2(S_1^2+S_2^2)=1$, $r^2=\pm 1$.
As in the case of IE, the quantity 

$$ Q={1 \over {4 \pi}} \int dx dy \vec S \cdot \left( \vec S_x 
\wedge \vec S_y \right ) \eqno(2) $$
may be called the topological charge. Both eq.(1) as well as IE in 
the (1+1) dimensional case reduce to one and the same equation - the 
well known (1+1) dimensional isotropic classical continuous Heisenberg 
ferromagnet model[3]. The Lax representation of eq.(1) was given in 
ref.[7] and some of its properties were studied in refs.[8-11].
The aim of this letter is to find the equivalent (geometrical 
and gauge) counterpart  namely the (2+1) dimensional nonlinear Schr\"odinger 
equation(NLSE) and to obtain physically interesting solutions such as line and 
curved solitons, exponentially localized and breaking solitons of eq.(1).

To begin with, let us find the geometrically equivalent counterpart 
of eq.(1), for $r^2=1$, that is, when $ \vec S^2 =S_1^2+ S_2^2+
S_3^2=1 $. For this purpose, we will extend the geometrical method 
applicable to (1+1) dimensional systems suitably to the (2+1) dimensional 
case. We associate a moving space curve parametrised by the arclength $x$, and endowed with an additional coordinate 
$y$ with the spin system[3,15,16]. Then the Serret-Frenet equation associated 
with the curve has the form

$$ \vec e_{ix} = \vec D \wedge \vec e_i,   \eqno (3a) $$
where

$$ \vec D = \tau \vec e_1 + \kappa \vec e_3    \eqno(3b)$$
and $ \vec e_i$'s, $i = 1,2,3 $ form the orthogonal trihedral. Mapping
the spin on the unit tangent vector

$$ \vec S(x,y,t) = \vec e_1 , \eqno(4) $$
the curvature and the torsion are given by

$$ \kappa (x,y,t) = ( \vec S_x^2)^{1\over2},  \eqno(5a)$$

$$\tau(x,y,t) = \kappa ^{-2} \vec S \cdot (\vec S_{x} \wedge \vec S_{xx}). 
\eqno(5b)$$
Due to the  orthonormality nature of the trihedral, $ \vec e_{it}. \vec e_i = 0$,
$ \vec e_{iy}. \vec e_i = 0$,  $i,j = 1,2,3$ and using the compatibility condition
$ \vec e_{ixy} = \vec e_{iyx}$, we find the equation for the $y$-part

$$ \vec e_{iy} = \vec {\gamma} \wedge \vec e_i, \eqno(6)$$ 
where $\vec {\gamma} =(\gamma_1,  \gamma_2, \gamma_3) $ and 

$$ \gamma_1 = u+\partial_x^{-1} \tau_y ,  \eqno(7a)$$

$$ \gamma _2 = - {u_x \over \kappa },     \eqno(7b)$$

$$\gamma_3= \partial_x^{-1}\left( \kappa _y-{\tau u_x \over \kappa}
\right ).   \eqno(7c)$$
Now, from eq.(1) and using eqs.(3) and (6), we can easily find the time 
evolution of the trihedral

$$ \vec e_{it} = \vec {\omega} \wedge \vec e_i, \eqno(8)$$  
with 

$$ \vec \Omega = (\omega_1, \omega_2, \omega_3) 
          = \left( {\kappa_{xy} \over \kappa }-\tau \partial_x^{-1} \tau_y,  -\kappa_y, 
          -\kappa \partial_x^{-1} \tau_y \right). \eqno(9)  $$

Ultimately, the compatibility condition  $\vec e_{ixt} =\vec e_{itx}$,  
which is also consistent with the relation $\vec e_{iyt} =\vec e_{ity}$, 
$i=1,2,3$ yields the following evolution equations for the curvature  
and torsion

$$ \kappa_t= -(\kappa \tau)_y-\kappa_x \partial_x^{-1} \tau_y  ,\eqno(10a)$$

$$ \tau_t= \left[ {\kappa_{xy}\over {\kappa}}- \tau \partial_x^{-1} \tau_y  
\right]_x + \kappa \kappa_y .  \eqno(10b)$$
On making the complex transformation[3],

$$ \psi (x,y,t) ={\kappa(x,y,t)\over 2} \exp \left [-i\int_ {-\infty }^{x}  
\tau(x',y, t) dx'\right ], \eqno(11)$$
the set of equations (10) reduce to the following (2+1) dimensional 
nonlinear Schr\"odinger equation(NLSE)

$$ i\psi_t = \psi_{xy}+ r^2V \psi, \eqno(12a)$$

$$ V_x = 2\partial_y{\mid \psi \mid }^2.     \eqno(12b)$$
Here, $r^2 = +1$, that is, we have the attractive type NLSE (The case $r^2 
= -1$ corresponds to the repulsive case). Eq.(12) was originally introduced by 
Zakharov[12] and was recently rederived by Strachan (for $r^2 = +1$)[13]. Its 
Painlev\'e property and some exact solutions were also obtained [14]. $N$- 
soliton solutions of eq.(12) for both the cases ($r^2=\pm 1$) can be found
in ref.[17]. Here, we have proved that eq.(12) is equivalent to eq.(1) in the 
geometrical sense.

Next, it is always of interest to note that eqs. (1) and (12) are also 
gauge equivalent in the sense of Zakharov and Takhtajan[18]. To this end, 
we write the Lax representation of eq.(1)[7], 

$$ \phi_{1x} = U_1 \phi_1 ,\eqno(13a)$$

$$\phi_{1t} = V_1 \phi_1 + \lambda \phi_{1y} , \eqno(13b)$$
where

$$ U_1 = {i \lambda \over 2 }S,\,\,\,\, S = \pmatrix{              
S_3 & rS^- \cr
rS^+ & -S_3
} , \eqno (14a)$$

$$V_1={\lambda \over 4}\left ([S,S_y]+2iuS\right),\,\,\,\,S^{\pm } = 
S_1 \pm iS_2 . \eqno (14b)$$
Here, $ \lambda $ is the eigen value parameter which satisfies the following
equations of Riemann wave type

$$  \lambda_t =  \lambda \lambda_y  .         \eqno(15)$$
It means that for solving eq.(1), we must in general use the non-isospectral 
IST. To obtain gauge equivalent counterpart of eq.(1), in the usual way
we consider the following gauge transformation

$$\phi_1 = g^{-1} \phi_2 , \eqno(16)$$
where $g(x,y,t)$ and $\phi_2(x,y,t,\lambda )$ are temporarily arbitrary matrix 
functions. Substituting eq.(16) into eq.(13), after some algebra  we get the 
following system of linear equations for $\phi_2$

$$ \phi_{2x}= U_2 \phi_2 , \eqno(17a)$$

$$\phi_{2t} = V_2 \phi_2 + \lambda\phi_{2y} \eqno(17b)$$
with

$$ U_2= {i\lambda\over2} \sigma_3+G,\,\,\,\,G= \pmatrix{           
 0 & \phi \cr 
 -r^2\phi & 0
 } , \eqno(18a)$$

$$V_2= i\sigma_3 \left({VI\over2}-G_y\right),\,\,\,\, I=diag(1,1), \eqno(18b)$$

$$ V= 2 \partial_x^{-1} \partial_y\left(\mid \psi\mid^2\right) \eqno(18c)$$
The compatibility condition of eq.(17) with (15) becomes (12), that is, eq.(1)
and eq.(12) are gauge equivalent to each other. The above transformation is in 
fact reversible and we can similarly prove that eq.(12) is gauge equivalent to 
eq.(1). 

The integrable eq.(1) allows an infinite number of integrals of motion.
The first two conservation laws are

$$\left( \vec S^2_x\right)_{t}+\frac14 \partial _x\left[ 
\vec S^2_x \partial _x^{-1}\left(  {
\vec S\cdot \vec S_x\wedge \vec S_{xx}\over \vec S^2_x}\right) _y\right] +$$ 
$$\frac14 \partial _y\left[ \vec S
\cdot \vec S_x\wedge \vec S_{xx}\right]=0,
\eqno(19a)$$ 

$$ \left[ \vec {S}\cdot \left ( \vec S_x \wedge \vec S_{xx} \right) \right]_t +
{1 \over 2} \partial _x \left[ { (\vec S^2_x)_x (\vec S^2_x )_y 
\over 4 } + \vec {S}\cdot \left ( \vec S_x \wedge \vec S_{xx} \right) \partial
_x^{-1} \left( \vec {S}\cdot \left ( \vec S_x \wedge \vec S_{xx} \right)\right)_y  
\right]$$
$$ + \partial_y \left\{ (\vec S^2_x)_x^2 + {{ 2 \left( \vec {S} \cdot
\left ( \vec S_x \wedge \vec S_{xx} \right) \right)^2} \over {\vec S^2_x}} - 
2 \left( \vec S^2_x \right)_{xx}-4 \vec S^4_x \right\} = 0 \eqno (19b)$$
and so on. Next, we present some important formulae which are just 
consequences of geometrical/gauge equivalence of eqs. (1) and (12). We have

$$ tr(S_x^2) = 8 \mid \psi \mid ^2= 2 \vec S_x^2 . \eqno(20a) $$ 
In a similar manner we find that

$$ -2i \vec S \cdot (\vec S_x \wedge \vec S_{xx})=tr(S_xSS_{xx})= 4(\psi \psi 
^*_x-\psi ^* \psi_x). \eqno(20b) $$
These relations are obviously equivalent to eq.(11). One may note that
these are of the same form as in the case of (1+1) dimensional Heisenberg 
chain [3]. 

Now, we wish to find a class of exact solutions of eq.(1),
such as line and curved solitons as well as exponentially localized solutions.
Introducing the stereographic variable

$$ S^+= S_1 +iS_2 ={2\omega \over {1+\mid \omega \mid^2}},\,\,\,\, S_3 
={1-\mid \omega \mid^2 \over {1+\mid \omega \mid^2}},  \eqno(21) $$
eq.(1) takes the form

$$ i\left( \omega_t-u\omega_x \right)+ \omega_{xy}-{2 \omega^* \omega_x 
\omega_y \over (1+\mid \omega \mid^2)} = 0, \eqno (22a) $$

$$ u_x + {2i( \omega_x \omega^*_y - \omega^*_x \omega_y) \over 
(1+\mid \omega \mid^2)^2} =0.  \eqno (22b)$$
On writing

$$ \omega = {g\over f}, \eqno(23) $$
where $g$ and $f$ are complex valued functions, and after using the 
Hirota's D-operators, eq.(22) becomes

$$ (iD_t-D_xD_y)(f^*\circ g) = 0 ,  \eqno(24a) $$        

$$ (iD_t-D_xD_y) \left( f^*\circ f-g^*\circ g \right) = 0, \eqno(24b) $$   

$$D_x \left( f^*\circ f+g^*\circ g \right) = 0,        \eqno(24c) $$   
while the potential $u(x,y,t)$ is 

$$ u(x,y,t) = -i{D_y\left( f^*\circ f+g^*\circ g \right) \over 
{f^*\circ f+g^*\circ g }}.        \eqno(24d) $$   
In terms of $g$ and $f$, the spin field $\vec S $ takes the form

$$ S^+= {2f^*g\over {\mid f \mid^2+\mid g \mid^2}},\,\,\,\,S_3=
{\mid f \mid^2-\mid g \mid^2 \over {\mid f \mid^2+\mid g \mid^2}}\,\,\,\, 
\eqno(25)  $$
and for $u(x,y,t)$ we get the following formula (using the properties 
of $D$-operators)

$$u_x(x,y,t) = -2i\partial^2_{xy} \ln{(\mid f \mid^2+\mid g \mid^2)}.
\eqno(26)  $$

The construction of the solutions to the M-I equation (1) now becomes 
standard. One expands the  functions $f$ and $g$ as a power series in the 
arbitrary parameter $ \epsilon $,

$$g= \sum_{n=0}^{\infty } \epsilon ^{2n+1}g_{2n+1},\,\,\,\, f=1+\sum_
{n=1}^{\infty } \epsilon^{2n}f_{2n}.    \eqno(27)  $$
Substituting these expansions into (24 a,b,c) and equating the coefficients of 
$\epsilon ^n $ yields

$$ \left[i\partial_t+\partial_x \partial_y\right] g_{2n+1}=-\sum_{k+m=n} 
D^{\prime}  (f^*_{2x}\circ g_{2m+1}),   \eqno(28a)  $$

$$\left[i\partial_t-\partial^2_{xy}\right](f^*_{2n}-f_{2n})= D^{\prime}\left(\sum _
{n_1+n_2=n-1} g^*_{2n_1+1}\circ g_{2n_2+1}-\sum _{m_1+m_2=n}f^*_{2m_1}\circ 
f_{2m_2} \right), \eqno(28b) $$

$$\partial_x(f^*_{2n}-f_{2n})= D_{x}\left(\sum _{n_1+n_2=n-1} g^*_{2n_1+1}\circ 
g_{2n_2+1}-\sum _{n_1+n_2=n}f^*_{2n_1}\circ f_{2n_2} \right), \eqno(28c) $$
with $D^{\prime}=iD_t-D_xD_y$,  $f_0=0$. In order to construct exact N-soliton 
(line and curved) solutions (N-SS) of eq.(1), we make the ansatz

$$ g_1= \sum_{j=1}^{N} \exp {\chi_j},\,\,\,\, \chi_j = \lambda_jx+m_j(y,t)+c_j.
\eqno (29)$$
We note here the important fact that $m_j(y,t)$ is an arbitrary 
complex function of $(y,t)$ of the form (see eq.(28a))

$$  m_j(y,t)= m_j(\rho ),\,\,\,\, \rho=y+i \lambda_j t,  \eqno(30)$$   
where $ \lambda _j $ is an arbitrary complex parameter. As an example, 
we write the forms of $g$ and $f$ for N=1 as 

$$g_1=\exp {\chi_1},\,\,\,\, f_2=\exp{2(\chi _{1R} +\psi)},\eqno(31)$$   
where 
$$\chi_1 = \chi_{1R}+i\chi _{1I}, \lambda_1=\eta +i \xi,
m_1=m_{1R}(\rho)+im_{1I}(\rho ), \chi_{1R}= \eta x+m_{1R}(\rho)+c_{1R},$$
$$ \chi_{1I}= \xi x +m_{1I}
(\rho)+c_{1I},\,\, c=ln(2 \eta /{\lambda_1^*}),\,\, \exp{2\psi}=
{-\lambda_1^2 \over{(\lambda_1+\lambda_1^*)^2}},$$
$$ m_{1R}(\rho )= 
Re m_1(\rho), m_{1I}(\rho )= Im m_1(\rho).
\eqno (32)$$
The corresponding 1-SS of eq.(1) takes the form

$$S_3(x,y,t)=1-{2 \eta^2  \over {\eta^2+\xi^2}} sech^2{\chi_{1R}}, \eqno(33a)
$$

$$S^+(x,y,t)={2\eta \over {\eta^2+\xi^2}} \left[i\xi - \eta tanh{\chi_{1R}}
\right] sech {\chi_{1R}}, \eqno(33b)$$ 
while for the potential

$$u(x,y,t)={2\eta \over {\eta^2+\xi^2}} \left(\xi m_{1R}'- \eta m_{1I}' 
\right) sech^2{\chi_{1R}} \eqno(33c)$$ 
in which the prime has been used to denote the differentiation with respect 
to the real part of the arguement. We note that the 1-SS(33) depends on two  
arbitrary functions $m_{1R}(\rho)$ and $m_{1I}(\rho)$  as in the case of some 
other (2+1) dimensional integrable equations. 

Now for the particular choice, 
         
         $$m_1=k_1(y+i\lambda_1t),$$ 
where $k_1$ is a complex constant, in eq.(33), 
we get the usual line 1-SS of eq.(1). For fixed $(y,t)$, it follows from (33) 
that  $\vec S \rightarrow (0,0,1)$ as $x \rightarrow \pm \infty$ and the 
wavefront  itself is defined by the equation $\chi_{1R}=\eta x+m_{1R}(\rho)+
c_{1R}=\eta x+k_1y-\xi t+c_{1R}=0$. For other choices of $m_1$, we can obtain 
more general solutions. Particularly, we present the dromion type localized 
solutions of eq.(1), the so-called induced localized structures/or induced 
dromions[14] for the potential $u(x,y,t)$.  This is possible by utilising the 
freedom in the choice of 
the arbitrary functions $m_{1R}$ and $m_{1I}$. For example, if we make the 
ansatz 

$$m_{1I}(\rho)= \kappa m_{1R}(\rho)= tanh(\rho_R),   
\eqno (34)$$

$$u=2 \eta (\xi -\eta \kappa) sech^2{\rho_R}sech[\eta x+ tanh{\rho_R}-
\eta x_0], \eqno (35)$$
where $\rho _R=y-\xi t$ and $k$ is a real constant. Similarly, the expressions 
for the spin can be obtained from eqs.(33). The solution (35) for $u(x,y,t)$ 
decays exponentially in all the directions, eventhough the spin $\vec S $ 
itself is not fully localized though bounded.
Analogously we can construct another type of ``induced dromion" solution
with the choice

$$m_{1I}= \kappa m_{1R} = \int {d \rho_R \over {(\rho_R+\rho_0)^2+1}}+m_0 ,
\eqno (36)  $$
where $\rho_0$ and $m_0$ are constants, so that

$$u(x,y,t)={2\eta (\xi - \kappa y)\over {(\rho_R+\rho_0)^2+1}} sech^2 \left [\eta x+
 \int {d \rho_R \over {(\rho_R+\rho_0)^2+1}}- \eta x_0\right]. \eqno (37)  $$
Generalizations of these solutions are also possible, which will be 
considered elsewhere.

Finally, we note here that we have a non-isospectral problem, as the 
spectral parameter $\lambda $ satisfies eq.(15). The above presented solutions
correspond to the constant solution of eq.(15), that is $\lambda =\lambda_1=$
constant. One may consider other interesting solutions of eq.(15). For example,
one can have a special solution

$$\lambda = \lambda_1= \eta(y,t)+i \xi (y,t) = {y+k+i\eta \over {b-t}}, \eqno 
(38)$$
where $b$, $k$ and $\eta$ are real constants. Corresponding to this case, we may call
the solutions of eqs. (1) and (12) as breaking solitons[19].
Using the Hirota method, one can also construct the breaking 1-SS of eq.(1)
associated with (38). For this purpose, we take $g_1$ in the form

$$g=g_1= \exp{\chi },\,\,\,\, \chi = ax+m+c =\chi_R+i\chi_I ,\eqno(39) $$ 
where $a=a(y,t)$, $m=m(y,t)$ and $c=c(t)$ are functions to be determined.
Substituting (39) into the first of eq.(28a), we get

$$ia_t+aa_y=0,\,\,\,\,im_t+am_y=0,\,\,\,iA_t+Aa_y=0, \eqno(40)$$
where $A=\exp(c)$. Particular solutions of eqs.(40) have the forms

$$ a=-i\lambda = {\eta -i(y+k) \over {b-t}},\,\,\,\, m=m{\left( y+k+i\eta 
\over {{b-t}} \right) },\,\,\,\, A={A_0\over {b-t}} , \eqno(41)$$ 
where $\eta $, $k$, $b$ and $A_0$ are some constants. From eqs. (28 b,c), 
we obtain

$$f_2 =B \exp {2\chi_R},\,\,\,\, B={\mid A_0 \mid^2 (y+k+i\eta )^2 \over 
{4{\eta }^2(b-t)^2}} .\eqno(42) $$

Now, we can write the breaking 1-SS of eq.(1) (using equations (25),
(39)-(42))

$$ S^+(x,y,t) = {2\eta \exp {i(\chi_I+ \phi)}(y+k-i\eta ) \over {\left[ (y+k)^2
+\eta ^2\right] ^{3 \over 2}}} {\left[ (y+k)coshz -i\eta sinhz\right] \over
{cosh^2z}} ,\eqno(43a) $$

$$ S_3(x,y,t)= 1-{2\eta^2 \over {[(y+k)^2+\eta^2]}} sech^2z , \eqno(43b)$$
where $z={\eta \over(b-t)}x-{1\over2} \ln{[(y+k)^2-{\eta }^2]}+\psi $, \,\,\,
$ \psi = \ln \mid  {A_0(y+k+i\eta ) \over {2\eta (b-t)}} \mid  $,\,\,\, 
$ \chi_I= -{(y+k) \over {(b-t)}}x+m_I \left( {y+k+i\eta \over {b-t}} \right) $.
We see that the solution (43) corresponds to an algebraically decaying solution
for large $x$, $y$.

Finally, we note that eq.(1) is a particular case of the following family 
of (2+1)  dimensional equation

$$ \vec S_t= \{ \vec S \wedge \vec S_y + u \vec S \}_x+ \vec F ,\eqno(44)$$
where $u$ and $\vec F $  satisfy the eq.(1b) and $ \vec S \cdot \vec F = 0$ 
respectively. Eq.(44) admits many integrable reductions, for example,
     
a) the isotropic M-I eq.(1), when $\vec F=0$;
     
b) the anisotropic M-I equation, when $ \vec F= \vec S \wedge A \vec S $,
where $A=diag(a_1, a_2, a_3)$, and 
     
c) the M-II equation, when $\vec F= mv \vec S_x+n \vec S_y$, where $v_x
=k(\vec S^2_x)_y$, $m$, $k$, $n$ are constants;\\
and so on. All of these equations are integrable in the sense that the 
corresponding Lax representations exist[7] and their gauge equivalent 
counterparts can be constructed[10](see also [20]-[21]). So further studies
of them will give more insight into the structure of nonlinear spin excitations 
in (2+1) dimensions.

\section* {Acknowledgements}

This work of M.L forms part of a Department of Science and Technology, 
Government of India sponsored research project. R.M. wishes to thank 
Bharathidasan University for hospitality during his visits to 
Tiruchirapalli. S.V. acknowledges the receipt of a Junior Research 
Fellowship from the Council of Scientific and Industrial Research, India.

\end{document}